\documentclass[aps,prl,twocolumn,showpacs,superscriptaddress]{revtex4}

\usepackage{graphicx}
\usepackage{graphicx}
\usepackage{epsfig}
\usepackage{amssymb}

\newcommand{\hf}{{_1\over^2}}
\newcommand{\ud}{\mathrm{d}\ }

\begin{document}
\title{Heat conductivity from molecular chaos hypothesis in locally confined billiard systems}
\author{Thomas Gilbert} 
\affiliation{Center for Nonlinear Phenomena and Complex Systems,\\
Universit\'e Libre de Bruxelles, Code Postal 231, Campus Plaine,
B-1050 Brussels, Belgium}
\author{Rapha\"el Lefevere}
\affiliation{Laboratoire de Probabilit\'es et Mod\`eles Al\'eatoires\\
UFR de Math\'ematiques Universit\'e Paris 7
 Case 7012, 75251 Paris Cedex 05, France }

\date{\today}

\begin{abstract}
We study the transport properties of a large class of locally confined
Hamiltonian systems, in which neighboring particles interact through hard 
core elastic collisions. When these collisions become rare and the systems
large, we derive a Boltzmann-like equation for the evolution of the
probability densities. We solve this equation in the linear regime and
compute the heat conductivity from a Green-Kubo formula. The validity of
our approach is demonstated by comparing our predictions to the results of
numerical simulations performed on a new class of high-dimensional
defocusing chaotic billiards.  
\end{abstract}


\maketitle

The understanding of Fourier's law and the computation of the heat
conductivity in Hamiltonian systems as a function of temperature and of the 
physical parameters remains to this day a challenging issue
\cite{cite}. In particular, establishing the necessary conditions the
dynamics must satisfy so as to justify a first principles based derivation
of Fourier's law has been the subject of ongoing discussions.

As a generic model of heat transfer in insulating crystalline
solids, one often considers a lattice of coupled particles with
nearest-neighbor interactions whose motion obeys Hamilton's equations. Thus
consider $N$ particles  of unit masses  located on a one-dimensional
lattice with positions and momenta 
$(\underline{\mathbf{q}}, \underline{\mathbf{p}}) \equiv 
\big\{(\mathbf{q}_i, \mathbf{p}_i)\big\}_{1\leq i\leq N}$, with
$\mathbf{q}_i, \mathbf{p}_i \in \mathbb{R}^d$. The Hamiltonian $H$ takes
the form
\begin{equation}
H(\underline{\mathbf{p}}, \underline{\mathbf{q}}) 
= \sum_{i=1}^N \left[\frac{p_i^2}{2} +V(\mathbf{q}_i)+
U(\mathbf{q}_{i}-\mathbf{q}_{i+1}) \right],
\label{Hamilton}
\end{equation}
where $V$ represents the interaction with the external substrate and $U$
the nearest-neighbor interactions \footnote{It is understood here that the
  positions $\mathbf{q}_i$ are measured with respect to a local referential
  at site $i$.}. 

After Peierls' work \cite{Peierls}, all attempts to give a satisfactory
derivation of Fourier's law in mechanical systems have focused on the study
of weakly anharmonic dynamics.  Using the Peierls-Boltzmann equation,
recent works have studied the effects of phonon collisions on the heat
conductivity \cite{BK,P,Spohn0,LS,ALS}. In this context, the
conductivity may be interpreted as a collision frequency between phonons.  

In this letter, we focus on the opposite limit, namely extremely anharmonic
interactions, and, under minimal assumptions on the chaotic nature of 
the dynamics, identify a class of models which display a universal
response to non-equilibrium thermal constraints. The motivation for this
study is twofold: 
First, the heat conductivity can be computed from first principles and
takes a simple form;
Second, as pointed out in \cite{GG08}, such systems of locally confined
particles in interaction find concrete applications in the study of
aerogels, materials in which gas particles are trapped in nano-size pores
and rarely interact among themselves. Assuming the validity of a
Boltzmann-like equation to describe such systems of rarely interacting
particles when they become large, we show that the heat conductivity of
such systems is generically equal to the average frequency of interaction
between the systems' components, \emph{i.e.} irrespective of the detailed
geometric properties of the confinement mechanism.  This will be checked in
detail by numerical simulations, showing the universality and power of the
Boltzmann approach to analyze the transfer of heat in the mechanical
systems we study.

To be specific, we consider the case of interaction potentials which take
only the values zero inside a region $\Omega_U\subset\mathbb{R}^d$ with
smooth boundary $\Lambda$ of dimension $d-1$, and infinity outside.
Likewise, the pinning potential $V$ is assumed to be zero inside a
bounded region $\Omega_V$ and infinity outside, implying that the motion of
a single particle remains confined for all times. The regions $\Omega_U$ and
$\Omega_V$ being specified, the dynamics is equivalent to a billiard in
higher dimension. An important quantity in such models is the average rate
of collisions between nearest-neighbors under equilibrium conditions. We
will be specifically concerned with the limit of rare collision events.

The shape of the region $\Omega_V$ determines the
nature of the local 
dynamics. In ref. \cite{GG08}, $\Omega_V$ was chosen to be a
semi-dispersive billiard with bounded horizon, thus ensuring strong chaotic
properties of the dynamics. In particular the fast decay of correlations of
the local dynamics was invoked to set up a stochastic equation
describing the energy exchange dynamics. It is our purpose to show that
this assumption can be relaxed: local ergodicity is enough to warrant the
identity between heat conductivity and frequency of energy exchanges. We
regard this as an important result which further validates the analogy
between this class of models and aerogels whose nanopores need not have
dispersing properties. 

\begin{figure}[thb]
\includegraphics[width = .47\textwidth]{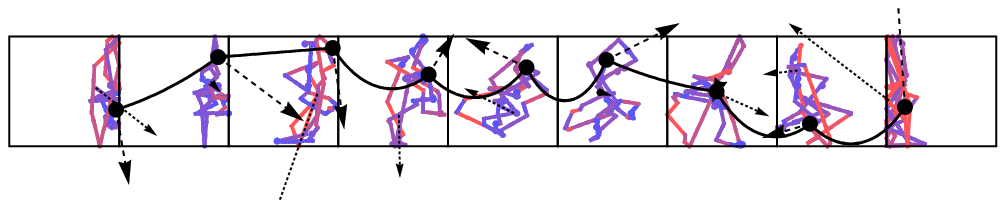} 
\includegraphics[width = .47\textwidth]{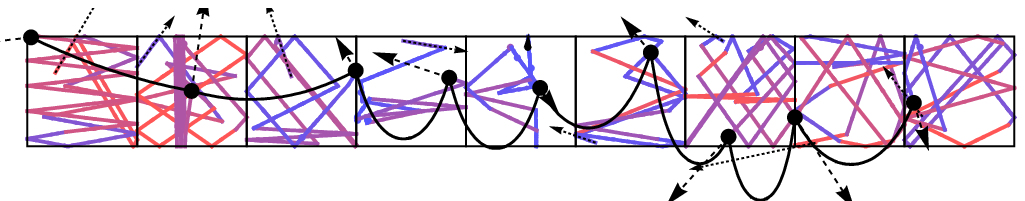} 
\includegraphics[width = .47\textwidth]{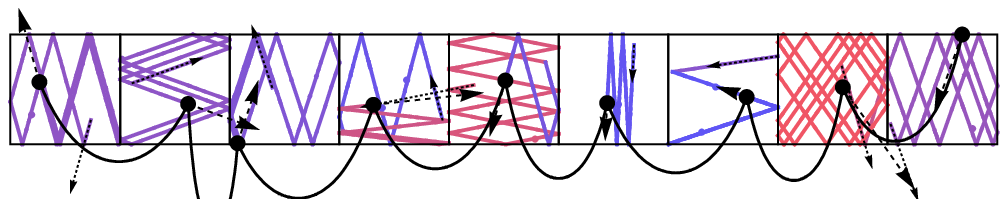}
\caption{(Color online) Typical trajectories of the square-strings model,
  displayed for increasing values of $a$, color-coded from blue to red
  according to their energies. Small and large arrows indicate
  initial and final velocities respectively. For large $a$, the system
  appears to be near-integrable, reflecting the rarity of interactions, but
  is nevertheless fully chaotic.}
\label{fig.square}
\end{figure}

Examples of the simplest type of billiards we may consider are periodic
arrays of square boxes in two dimensions in each of which a single hard disk
particle moves freely, but can still perform collisions with neighboring
disks by interacting through the confining walls, for instance, provided we
let the cells overlap a bit. The specific nature of the
interaction mechanism at play is however not relevant in our formalism. We
will instead consider point particles moving freely in two-dimensional
square boxes of unit sides and 
interacting among nearest neighbors when the Euclidean 
distance between them becomes equal to a parameter which we denote by $a$.
At that point, they exchange their longitudinal velocities, \emph{i.e.} 
the velocity components in the direction of their relative motion.
We refer to this model as the \emph{square-strings model}.
The interaction may be depicted by attaching 
strings of lengths $a$ separating neighboring particles, as shown in
Fig.~\ref{fig.square}.
In this case, we take $\Omega_V=[-1/2,1/2]^2$, 
and $\Omega_U = \mathbb{D}^2_{(-1,0)}(a)$, the disk of radius $a$ with
center at $(-1,0)$ \footnote{The origin of the disk is shifted because the
  positions of the particles are measured with respect to the center of the
  cell $\Omega_V$.}.
We note that, in the absence of interactions, the dynamics of the
individual particles is pseudo-integrable; it is ergodic on the
configuration space for most values of the velocity directions, but is
known to be non-mixing. We will consider this model in some details below
and provide numerical evidence that the analysis which follows applies to
it.

This model can be compared to a simpler class of \emph{complete exchange
  models}, specified by square well potentials, obtained from
Eq.~(\ref{Hamilton}) with $d=1$, as a limit of models with smooth
interaction potentials.  In this case, $\Omega_V=[-b,b]$ and
$\Omega_U=[-a,a]$.  Each particle on the lattice moves freely on a
one-dimensional cell of size $2 b$, changing directions at the
boundaries. The interaction  between a pair of particles acts when the
difference between the positions of the two particles reaches the value
$a$, at which point they exchange their velocities.

In the general $d$-dimensional set-up,  the particles move freely inside
their respective cells, bouncing off the walls elastically, until the
vector $\mathbf{q}_i - \mathbf{q}_{i+1}$ (resp. $\mathbf{q}_{i-1} -
\mathbf{q}_{i}$) reaches the boundary $\Lambda$. The corresponding
particles then exchange the components of their velocities in
the direction normal to the boundary $\Lambda$, {\em i.e.} longitudinal to
the direction of their relative motion. 

The Hamiltonian (\ref{Hamilton}) may be written as a sum of local terms, 
$h_i(\underline{\mathbf{p}}, \underline{\mathbf{q}}) 
= p_i^2/2 + V(\mathbf{q}_i)
+ 1/2 \left[U(\mathbf{q}_{i-1} - \mathbf{q}_{i}) +
U(\mathbf{q}_{i} - \mathbf{q}_{i+1})\right]. $
This allows one to define a function describing the local transfer of
energy by computing the variation in time of the local energy $h_i$ along
the solutions of the equations of motion,
$\frac{d}{dt}h_i(\underline{\mathbf{p}}, \underline{\mathbf{q}}) = j_{i-1}
- j_i$, where the local energy current between sites $i$ and $i+1$ is
defined as,  
$j_i \equiv \hf (\mathbf{p}_i + \mathbf{p}_{i+1})\cdot\nabla 
U(\mathbf{q}_i - \mathbf{q}_{i+1})$, which, for hard core interactions,
becomes 
\begin{equation}
j_i = -\hf \delta_{\Lambda}(\mathbf{q}_i - \mathbf{q}_{i+1})
|p^\bot_{i+1}-p^\bot_{i}|^+
\left[(p^\bot_{i+1})^2- (p^\bot_{i})^2\right], 
\label{gencurrent}
\end{equation}
where $|x|^+=x$, if $x\geq 0$, and $0$ otherwise, $p^\bot_{i}=
\mathbf{p}_i\cdot \widehat{\mathbf{n}}$ is the component of 
the vector $\mathbf{p}_i$ in the direction of the unit vector
$\widehat{\mathbf{n}}$, normal to the boundary $\Lambda$, and
$\delta_{\Lambda}$ denotes the delta 
function concentrated on this boundary. The first factor corresponds to the
localization of the collisions in configuration space, the second one gives
the rate at which collisions occur and the last one corresponds to the
exchange of longitudinal components of the kinetic energies.  

Starting from the Liouville equation for the evolution of probability
densities on phase space, it is straightforward to derive an equation for
the evolution of the probability density of a single particle in a given
cell.  It involves the probability distribution of the pairs of particles
which consist of the particle itself and either of its nearest-neighbors on
the lattice.  The Boltzmann approximation simply amounts to assuming that
this two-particle distribution factorizes in terms of the one-particle
distributions $f_i$ at each site.  To justify this assumption, one needs to
show that a version of molecular chaos holds in our models. Namely, that
the dynamical variables involved in the successive collisions between two 
neighbors are independent at the times of collisions. For that purpose, we
require two ingredients:
First, local correlations are typically destroyed after a collision between
neighboring particles; 
Second, the number of particles must be very large, so
that in the long run, the whole system plays the role of a reservoir for
the specified pair of nearest neighbors. How these conditions are
realized in the models we consider and, in particular, in the
square-strings model which we test numerically, is not yet fully
elucidated. We interpret the first condition as requiring interactions
to be rare compared to the collisions within a single cell. In the 
square-strings model, it amounts to taking the maximal separation close to
the length of the diagonal joining opposite corners of  neighboring boxes
($\equiv\sqrt{5}$), as in the third panel of Fig.~\ref{fig.square}. 

We denote by  $\underline{f}= \big\{f_i(\mathbf{p}, \mathbf{q},
t)\big\}_{1\leq i\leq N}$, the set  of the marginal probability
distributions of each particle in each cell. The Boltzmann equation for
this set of probability densities is
\begin{equation}
\frac{d}{dt}f_i(\mathbf{p}, \mathbf{q}, t) =
-\mathbf{p}\cdot\nabla_\mathbf{q} f_i + L^\mathbf{w}f_i + L^\mathrm{c}_{i,i+1} 
(\underline{f}) + L^\mathrm{c}_{i, i-1} (\underline{f}).
\label{Bol}
\end{equation}
Here $L^\mathrm{w}$ accounts for the collisions of the particles with the
walls of  
their respective cells, and $L^\mathrm{c}_{i,i\pm1}$ for the collisions of
the $i$-th particle with the $i\pm1$th, {\em viz.}  
\begin{eqnarray}
  \lefteqn{L^\mathrm{c}_{i,i\pm1}(\underline{f}) 
    = \int \ud \mathbf{p}_a \ud \mathbf{q}'
    \delta_{\Lambda}(\mathbf{q} - \mathbf{q}')
    |p^\bot-p_a^\bot|^+}\label{Lc}\\ 
  &&
  \times [f_{i\pm1}(\mathbf{p}_b, \mathbf{q}')
  f_i(\mathbf{p}_c, \mathbf{q}) - f_i(\mathbf{p}, \mathbf{q})
  f_{i\pm1}(\mathbf{p}_a, \mathbf{q}')],\nonumber
\end{eqnarray}
with $p_b^\bot=p^\bot$, $p_c^\bot=p_a^\bot$, $\mathbf{p}_c - p_c^\bot 
\widehat{\mathbf{n}} = \mathbf{p} - p^\bot \widehat{\mathbf{n}}$, 
and $\mathbf{p}_b - p_b^\bot \widehat{\mathbf{n}} = \mathbf{p}_a - p_a^\bot
\widehat{\mathbf{n}}$.  
One can check that the distribution 
\begin{equation}
  \mu_\mathrm{eq} \equiv
  \prod_{i=1}^N f_i(\mathbf{p}_i, \mathbf{q}_i) = Z^{-1}
  \prod_{i=1}^Ne^{-\beta p^2_i/2}{\bf 1}_{\Omega_V}(\mathbf{q}_i) 
  \label{mueq}
\end{equation}
is stationary for any inverse temperature $\beta$.  Applied to this
distribution, the advection term in Eq. (\ref{Bol}) is zero except on the
cell borders where it cancels  with $L^\mathrm{w} f_i$.
$\beta$ may be fixed by imposing identical thermal boundary conditions at
both ends of the lattice. 

When the system is set out of equilibrium by
imposing different temperatures at its boundaries, we proceed
with a standard Chapman-Enskog expansion around a local equilibrium distribution,
\begin{equation}
  \mu_\mathrm{leq} \equiv
  \prod_{k=1}^Nf_k(\mathbf{p}_k, \mathbf{q}_k) = 
  Z^{-1}\prod_{k=1}^Ne^{-\beta_k p^2_k/2}{\bf 1}_{\Omega_V}(\mathbf{q}_k) ,
  \label{loceq}
\end{equation}
with $\beta_k=\hat\beta(k/N)$ for some smooth function $\hat\beta$,  taking
as a small parameter the local temperature gradient. Plugging Eq.~(\ref{loceq})
into (\ref{Bol}), we observe that only terms of second-order in the
temperature gradient survive. This is in contrast with the case 
of an ordinary gas of colliding particles.  This simplification occurs
because the advection term of the Boltzmann equation (\ref{Bol}) acts only
on the position variable within each cell and therefore not as a gradient
on the lattice dependent variables.
This means that local averages with respect to the distribution  (\ref{loceq})
are identical to local averages with respect to the true non-equilibrium
stationary state, denoted $\langle \cdot\rangle_{\mathrm{neq}}$, up to
$1/N^2$ corrections.  

In particular, one may compute the average current
(\ref{gencurrent}) with respect to the measure (\ref{loceq}) and get (with
$\beta_i=T^{-1}_i$),  
\begin{equation}
\langle j_i\rangle_\mathrm{neq}= - \nu(T_i)(T_{i+1}-T_i)+ \mathcal{O}(1/N^2),
\end{equation}
where,
$
\nu(T_i)= \langle
\delta_{\Lambda}(q_i-q_{i+1})|p^\bot_i-p^\bot_{i+1}|^+\rangle_{T_i} 
$
is readily interpreted as the average collision frequency between the
neighbors $i$ and $i+1$, with respect to a global equilibrium measure at
temperature $T_i$, Eq.~(\ref{mueq}).  This computation therefore shows that
the conductivity $\kappa(T_i)$, defined as 
\begin{equation}
  \kappa(T_i)\equiv \lim_{N\rightarrow\infty}
  - \frac{\langle j_i\rangle_\mathrm{neq}}
  {T_{i+1}-T_{i}}, 
\label{conductivity2}
\end{equation}
is identical to $\nu(T_i)$. Furthermore a simple scaling argument shows that
$\kappa(T_i) = \nu(T_i) = \sqrt{T_i} \nu,$ where $\nu$ denotes the collision
frequency computed at unit temperature.  Being the result of an equilibrium
integration, the frequency may be computed with arbitrary precision. 

In order to get a better picture of the process that is described by the
Boltzmann equation (\ref{Bol}), we linearize the equation around the global
equilibrium solution (\ref{mueq}).  Doing so, we obtain an equation similar
to (\ref{Bol}), but with the collision operators 
$L^\mathrm{c}_{i,i\pm1}$ now replaced by $L^\mathrm{lin}_i$, 
\begin{equation}
\frac{d}{dt}f_i(\mathbf{p}, \mathbf{q}, t) =
-\mathbf{p}\cdot\nabla_\mathbf{q} f_i + L^\mathrm{w}f_i + 
2L^\mathrm{lin}_i f, 
\label{Boll}
\end{equation}
where the linearized collision operator $L^\mathrm{lin}_i$ is obtained from
Eq.~(\ref{Lc}) by replacing $f_{i\pm1}$ by equilibrium distributions at
common inverse temperature $\beta$. 

The interpretation of the stochastic process described by the linearized
collision operator is straightforward: when collisions take place, the
particles velocities are updated as though they collided with stochastic
thermal walls at inverse temperature $\beta$ \cite{LSb}. At each collision,
the new velocities are  independent from the previous ones.  

With this prescription, 
we now compute the conductivity using the Green-Kubo formula, which is
derived as follows. Integrated over time, the energy current between sites
$i$ and $i+1$ takes the form 
\begin{equation}
  J_i([0,t])=\int_0^t j_i(s) ds = \hf 
  \sum_{0\leq s^k_i\leq t} \left[ p_{i}^\bot(s^k_i)^2 -
    p^\bot_{i+1}(s^k_i)^2 \right]\,,
\label{timeintegrate}
\end{equation}
where the $(s^k_i)_{k\in\mathbb{N}}$ are the successive collision times
between particles $i$ and $i+1$.
The Green-Kubo formula, reads in our case,
\begin{equation}
\kappa_\mathrm{GK}(T)=\frac{1}{2 NT^2}\lim_{t\rightarrow\infty}
\frac{1}{t}\sum_{i,k=1}^N\Big\langle J_i([0,t])J_k([0,t])\Big\rangle_{T}. 
\label{GK1}
\end{equation}
Using the expression (\ref{timeintegrate}), (\ref{gencurrent}),
translation-invariance and the independence of the transfer of energy at
each collision, we get, after some calculations, 
\begin{eqnarray}
\lefteqn{\kappa_\mathrm{GK}(T) =}\label{kappaeqnu}\\
&& \frac{1}{8 T^2} \left\langle
\delta_{\Lambda} (\mathbf{q}_0 - \mathbf{q}_{1}) |p^\bot_0 - p^\bot_{1}|^+
\left[(p^\bot_0)^2 - (p^\bot_{1})^2\right]^2 \right\rangle_{T},\nonumber
\end{eqnarray}
which, after further computations turns out to be equal to the collision
frequency, $\kappa_\mathrm{GK}(T) = \nu(T)$. 

The square-strings model displayed in Fig. \ref{fig.square} lends itself
to a detailed study of the dependence of the ratio $\kappa/\nu$ on the
parameter values $a$. 

\begin{figure}[thb]
  \includegraphics[width = .45\textwidth]{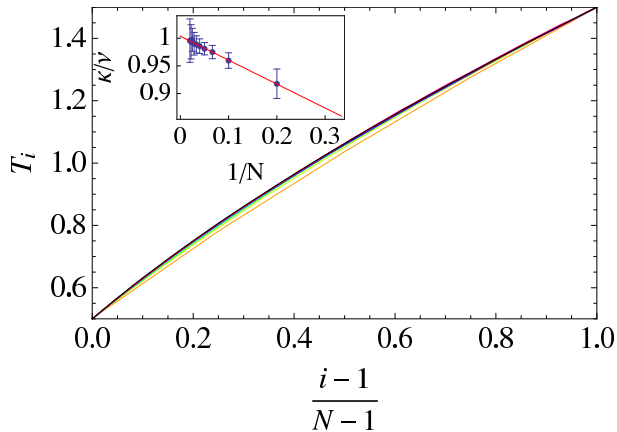}
  \caption{(Color online) Non-equilibrium temperature profiles of the
    square-strings model with $a=2.08$, for increasing values of $N = 5,
    10,\dots, 50$. The black curve is the stationary solution of the heat
    equation. The inset displays the corresponding measurements of
    $\langle\kappa(T_i)/\nu(T_i)\rangle$. The infinite 
    $N$ extrapolation, $\kappa/\nu = 1.0037$, is the approximate 
    heat conductivity reported in Table \ref{tab.kappanu}.} 
  \label{fig.profile}
\end{figure}

To this end we consider systems of varying sizes $N$
with both ends in contact with stochastic thermal baths at respective
temperatures $T_- = 1/2$ and $T_+ = 3/2$. This gives rise to
non-equilibrium stationary states with temperature profiles such as
displayed in Fig. \ref{fig.profile}, which, as $N$ increases, approach the
corresponding solution of the heat equation, $\partial_x[\kappa(T(x)) \partial_x
T(x)] = 0$, with $\kappa(T(x))\propto\sqrt{T(x)}$.  
The ratio $\kappa/\nu$ is obtained by linearly extrapolating to
$N\to\infty$ finite $N$ measurements of the spatial averages of
$\kappa(T_i)/\nu(T_i)$, with 
$\kappa(T_i)$ defined by Eq. (\ref{conductivity2}) and $\nu(T_i)$ the
collision frequency at the local temperature, as functions of $1/N$.

These values are reported in Table \ref{tab.kappanu}. Notice the excellent
agreement with the prediction $\kappa=\nu$, Eq.~(\ref{kappaeqnu}), as the
value of the parameter $a$ gets closer to its maximal allowed value, the
limit of rare collisions, in close agreement with the results presented in
\cite{GG08} for a class of coupled semi-dispersing billiards. In particular
we underline that the parameter range of validity of our result is very
similar to that observed in \cite{GG08}, which further validates that it is
independent of the detailed nature of the local dynamics.
\begin{table}[htb]
\vspace{.1cm}
\begin{tabular}{||c|c||c|c||c|c||}
\hline
$\sqrt{5}-a$&$\kappa/\nu$&$\sqrt{5}-a$&$\kappa/\nu$
&$\sqrt{5}-a$&$\kappa/\nu$
\\
\hline
1.029 & 6.000 & 0.618 & 1.892 & 0.322 & 1.320 \\
\hline
0.870 & 3.159 & 0.511 & 1.632& 0.236 & 1.0718\\
\hline
0.736 & 2.336 & 0.413 & 1.452 & 0.155 & 1.0037\\
\hline
\end{tabular}
\caption{Measurements of $\kappa/\nu$ for selected values of $a$, obtained
  from data similar to Fig. \ref{fig.profile}. Our
  results indicate that $\kappa/\nu\to 1$ as $a\to\sqrt{5}$, in agreement
  with Eq.~(\ref{kappaeqnu}).} 
\label{tab.kappanu}
\end{table}

To summarize, we have showed that the derivation of Fourier's law in a
large class of locally confined particle systems with hard-core
interactions can be achieved from a Boltzmann-type approach with the main
result that, in the appropriate limits, the heat conductivity is identified
with the collision frequency. 

The same identity was derived in \cite{GG08} in the context of
semi-dispersing billiards. The comparison is interesting since, in
contrast, chaos in the square-strings model 
results from a defocusing mechanism which takes place after particles 
interact. The identity between conductivity and collision frequency
therefore proves to be more general as it accounts for the transport
properties of systems lacking the local mixing property. In fact, the only
dynamical property which is a priori necessary in our derivation is 
ergodicity of the local dynamics, {\em  i.e.} in the absence of
interactions. This property guarantees that two neighbors always interact
provided the coupling is switched on, and that the fraction of time during
which  they interact is proportional to a fixed geometrical factor which
can be adjusted by tuning the systems' parameters. 

The square-strings model is a perfect example of a system which lends
itself with ease to a precise and reliable numerical analysis, while
retaining the molecular chaos property. The square-strings model is
actually a kind of higher dimensional fully chaotic stadium and displays a
very rich structure of dynamical properties.  

We regard the proof of the molecular chaos hypothesis upon which our
computation relies as a promising and realistic way to eventually
obtain a clear picture of the different mechanisms responsible for the
origin of Fourier's law in a large class of mechanical systems.

\begin{acknowledgments}
The authors thank J. Bricmont, M. D. Jara Valenzuela, P. Gaspard,
A. Schenkel, and L. Zambotti for useful discussions.  TG is financially
supported by the 
Fonds de la Recherche Scientifique F.R.S.-FNRS and has additional
support from the Belgian Federal 
Government  IAP project ``NOSY''. RL acknowledges financial support from
ANR network LHMSHE.
\end{acknowledgments}


\begin{thebibliography}{10}

\bibitem{cite} A. Dhar, Adv. Phys. \emph{in press} (2008); arXiv:0808.3256.

\bibitem{Peierls} 
  R. Peierls,
  Ann. Phys. (Ger.) {\bf 3}, 1055-1101 (1929). 

\bibitem{BK} 
  J. Bricmont and A. Kupiainen, 
  Com. Math. Phys. {\bf 274} 555-626 (2007).

\bibitem{P} 
  A. Pereverzev,
  Phys. Rev. E {\bf 68} 056124 (2003). 

\bibitem{Spohn0}  
  H. Spohn, 
  J.  Stat. Phys. {\bf 124} 1041-1104 (2006). 

\bibitem{LS} 
  R. Lefevere and A.Schenkel,
  J. Stat. Mech. L02001 (2006). 

\bibitem{ALS}
  K.~Aoki, J.~Lukkarinen, H.~Spohn, 
     J. Stat. Phys. {\bf 124} 1105-1129 (2006).  


\bibitem{GG08}
  P.~Gaspard and T.~Gilbert, 
  Phys. Rev. Lett. {\bf 101} 020601 (2008); New J. Phys. \emph{in press}
  (2008).

\bibitem{LSb}
  J. L. Lebowitz and H. Spohn, 
  J. Stat. Phys. {\bf 19}, 633 (1978).


\end{thebibliography}
\end{document}